\documentclass[a4paper,
               keeplastbox,   
               ]{jacow}
\usepackage{pdfpages,multirow,ragged2e} %
\makeatletter%
	\ifboolexpr{bool{xetex}}
	 {\renewcommand{\Gin@extensions}{.pdf,%
	                    .png,.jpg,.bmp,.pict,.tif,.psd,.mac,.sga,.tga,.gif,%
	                    .eps,.ps,%
	                    }}{}
\makeatother

\ifboolexpr{bool{xetex} or bool{luatex}} 
 {}                                      
 {\usepackage[utf8]{inputenc}}           

\usepackage[USenglish]{babel}

\begin{document}

\title{A Method for Passive Streaker LPS Reconstruction\thanks{This work was supported by operation funds of the European XFEL.}}

\author{Sergey Tomin\thanks{sergey.tomin@desy.de}, Igor  Zagorodnov, \\
Deutsches Elektronen-Synchrotron DESY, Notkestr. 85, 22607 Hamburg, Germany}
	
\maketitle

\begin{abstract}
Understanding the electron beam distribution in the longitudinal phase space (LPS) is crucial for free electron laser (FEL) facilities. Conventionally, LPS diagnostics utilize radio frequency (RF) deflecting structures to streak the electron beam transversely, mapping the longitudinal bunch distribution onto a transverse plane for observation. However, RF structures are complex and costly, especially for high-energy machines like the European XFEL. Wakefield structures have emerged as a promising alternative, offering simplicity in construction and minimal maintenance costs. However, they suffer from nonlinear streaking, requiring image reconstruction for LPS distribution. Several iterative algorithms have been developed for LPS reconstruction using passive wakefield streakers in recent years. This paper proposes a simple, computationally efficient method tailored for cases with known beam current profiles.
\end{abstract}

\section{Introduction}\label{sec:1}

\begin{figure*}[!tbh]
    \centering
    \includegraphics*[width=0.9\textwidth]{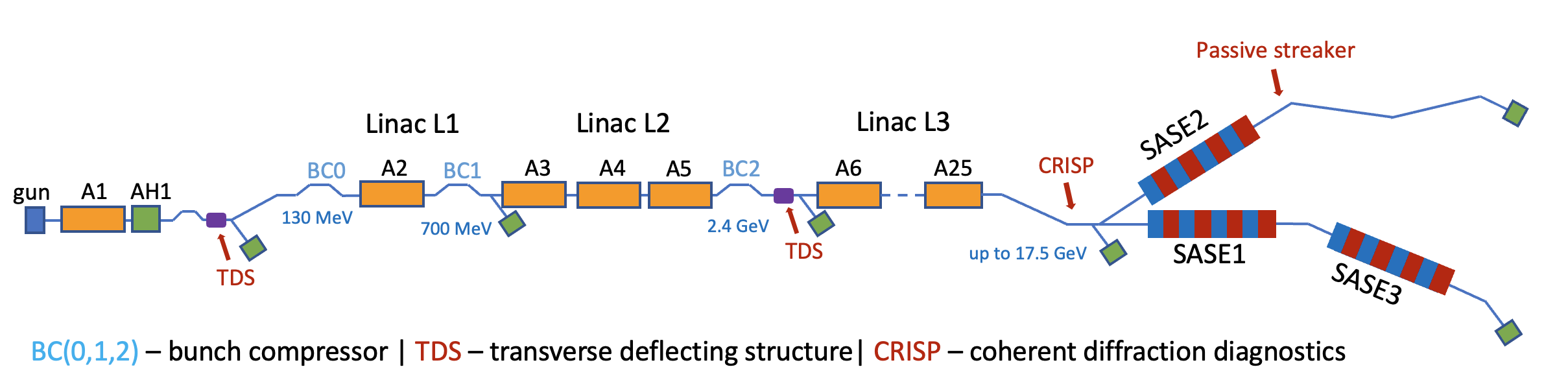}
    \caption{The European XFEL layout.}
    \label{fig:layout}
\end{figure*}

Longitudinal phase space (LPS) diagnostics are crucial for evaluating beam quality in free electron laser (FEL) facilities. The conventional method uses a radio frequency transverse deflection structure (TDS) \cite{TDS, XTCAV2}. However, due to the complexity and high cost of RF deflectors, especially at higher beam energies, the European XFEL, operating up to 17.5 GeV \cite{XFEL}, does not use TDS after the main linac L3, as shown in Fig.~\ref{fig:layout}.
A corrugated structure, consisting of a small-radius corrugated pipe or adjustable-gap metal plates, was initially proposed to remove linear energy correlation (chirp) in relativistic electron beams \cite{Bane_2012}, with experimental confirmation in \cite{Emma2014}. When an electron beam passes near the corrugated wall, it experiences a time-correlated transverse kick. In 2015, the use of a corrugated structure as a passive streaker was proposed \cite{novo2015}, though its transverse kick is nonlinear, requiring reconstruction of the longitudinal bunch distribution \cite{bane_zagor2016}.

The experimental demonstration of the wakefield passive streaker for measuring beam current profiles was presented in \cite{Bettoni2016, Seok2018}, along with proposed reconstruction methods. Recently, a passive streaker was employed for LPS measurement, and a new reconstruction algorithm was introduced \cite{PSI2022}. Inspired by SwissXFEL results, the European XFEL installed a corrugated structure after the SASE2 undulator, see Fig.~\ref{fig:layout}, implementing the same reconstruction method \cite{tomin_ipac, PSI2022, dijkstal2024}.

The reconstruction algorithm \cite{Bettoni2016, Seok2018, PSI2022, dijkstal2024} requires precise knowledge of both transverse and longitudinal components of the wake potential. We propose a fast LPS reconstruction algorithm that bypasses the need for transverse wake potential knowledge but requires the current profile, enabling online reconstruction in the control room at 1 Hz or faster.

\section{Reconstruction algorithm}\label{sec2}

\begin{figure}[!htb]
   \centering
   \includegraphics*[width=.9\columnwidth]{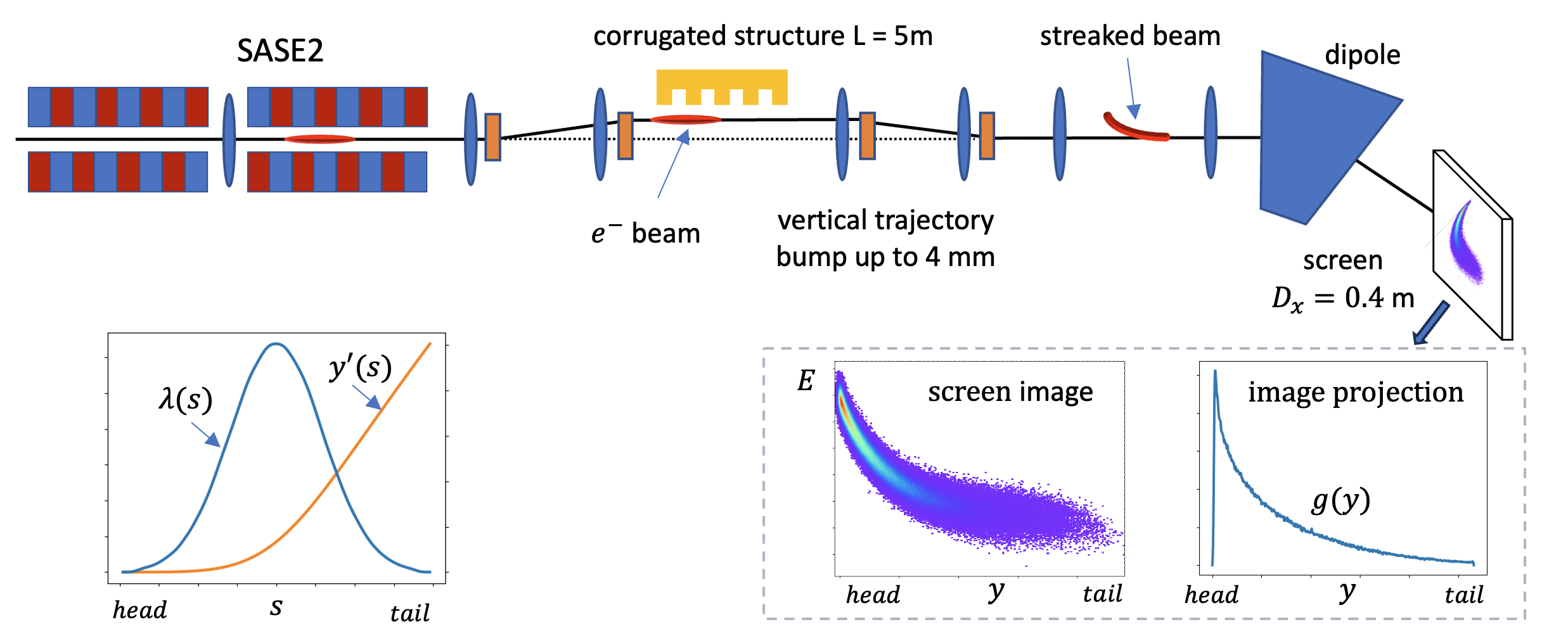}
   \caption{Simplified layout of the LPS diagnostics with a corrugated structure. The bottom row shows the passive streaker’s working principle: on the left, a Gaussian current profile and nonlinear kick; on the right, the streaked beam image on the screen with its horizontal projection $y$.}
   
   \label{fig:streaker}
\end{figure}

The longitudinal phase space diagnostic, based on a passive streaker, is located downstream of the SASE2 undulator. It uses a fixed 5-meter corrugated structure to streak the beam vertically in $y$. The transverse kick strength varies with the distance between the beam and the corrugated plate, controlled by a vertical trajectory bump (Fig. \ref{fig:streaker}). The screen is placed in a dispersive section with a horizontal dispersion of $D_x=0.4$ m. Further details are in \cite{tomin_ipac, dijkstal2024}.
The particle's position on the screen is depends on the transverse kick $y'(s)$ it receives from the passive streaker and the $R_{34}$ element of the transfer matrix $R$ from the streaker to the screen. The relationship in linear approximation is:
\begin{equation}\label{eq:map}
y(s) = y_0 + R_{34} y'(s).
\end{equation}
where $s$ represents the particle's longitudinal coordinate within the electron bunch, $y_0$ is initial vertical coordinate of the particle.

The magnitude of the deflection is given by:
\begin{subequations}
\label{eq:kick}
\begin{align}
y'(s) =& \frac{e V(s)}{c p_s}\approx \frac{e V(s)}{E_0} \label{eq:kick1} \\
V(s) = -&\left(\frac{Z_0 c}{4 \pi}\right)Q L W_{\lambda d}(s) \label{eq:kick2}
\end{align}
\end{subequations}
where $V(s)$ is the voltage induced by wakefields, $p_s$ the particle's longitudinal momentum, $E_0$ the beam energy, $Q$ the charge, $L$ the streaker length, and $Z_0=377 \Omega$ the characteristic impedance of free space. The wake potential $W_{\lambda d}(s)$ is the convolution of the dipole wake $\textit{w}_d(s)$ from the corrugated plate \cite{bane_zagor2016} and the normalized bunch shape $\lambda(s)$:
\begin{equation}\label{eq:conv}
W_{\lambda d}(s) = \int_{0}^{\infty} ds' \textit{w}_{d}(s) \lambda(s - s').
\end{equation}

Besides the dipole component of the wake, \(\textit{w}_{d}(s)\), the wake from the corrugated plate also includes quadrupole \(\text{w}_q(s)\) and longitudinal components \(\text{w}_l(s)\) \cite{bane_zagor2016}. 

The transverse kick is nonlinear and varies with the bunch current profile, leading to a nonlinear transformation between the longitudinal coordinate $s$ and the screen coordinate $y$, as shown in Fig.~\ref{fig:streaker}. Reconstruction methods iteratively determine the current profile and wake to define this transformation \cite{Bettoni2016, Seok2018, PSI2022}.

At the European XFEL, the current profile is continuously monitored based on the spectroscopy of coherent diffraction radiation (CDR). The THz and infrared spectrum of the CDR is captured by a multistage grating spectrometer named CRISP \cite{CRISPwesch}, which employs a reconstruction method to ascertain the current profile \cite{CRISP, CRISPrec}. The CRISP current profile reconstruction has a number of limitations that may affect the accuracy of the LPS reconstruction. These will be discussed in Discussion Section.

If the current profile is known, the transformation $s(y)$ can be deduced using the following approach. Let's consider the normalized current profile $\lambda(s)$ and normalized screen projection $g(y)$, both starting from zero, such that:
\begin{equation}
\int_{0}^{s_{max}}\lambda(s) ds = \int_{0}^{y_{max}} g(y) dy = 1
\end{equation}

\begin{figure}[!htb]
   \centering
   \includegraphics*[width=.9\columnwidth]{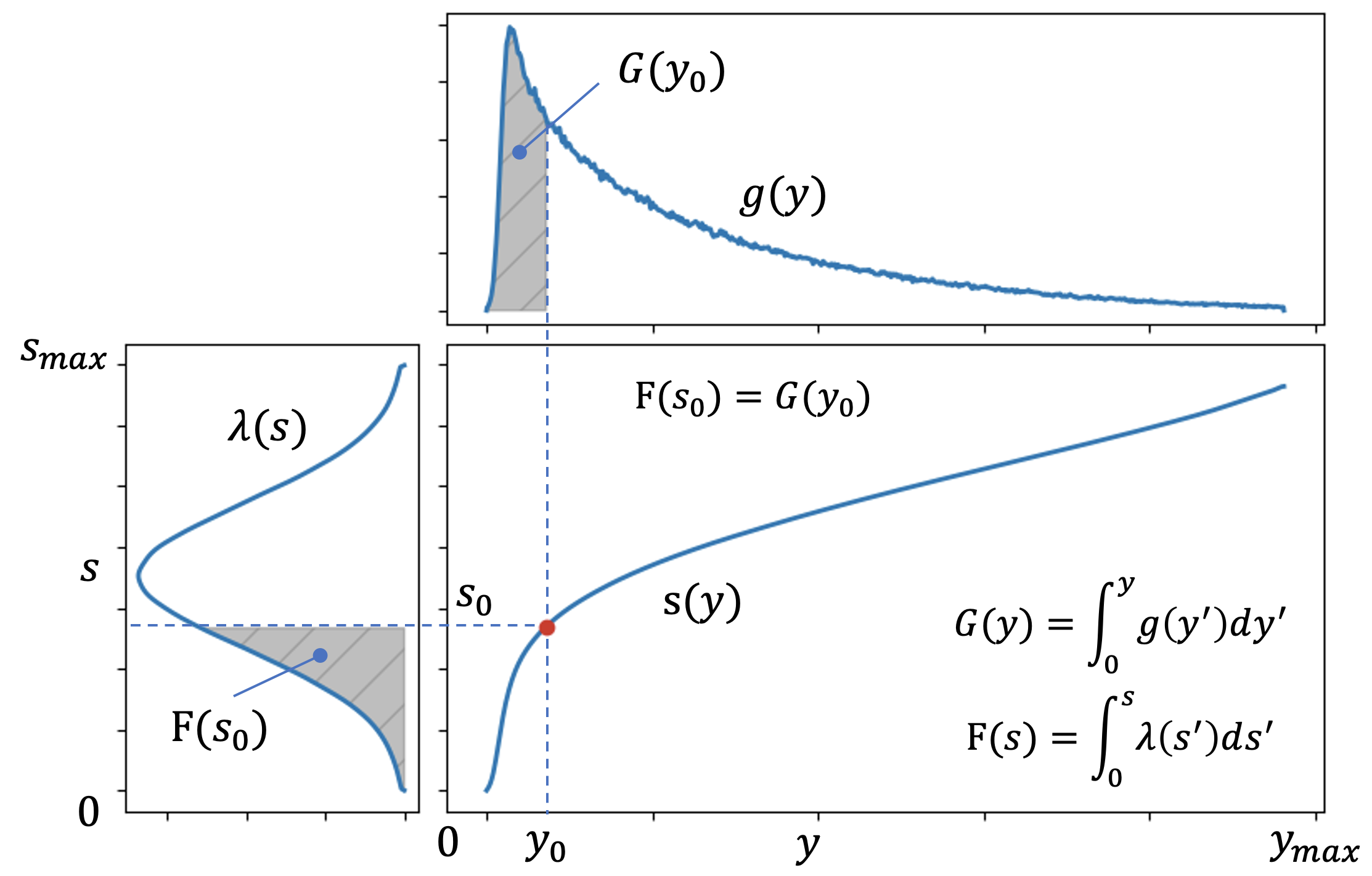}
   \caption{Illustration of the method for determining the transformation $s(y)$.}
   \label{fig:explained}
\end{figure}

In this context, the fraction of the beam charge from $0$ to $s_0$, or area $F(s_0)$,  should correspond to an equal area $G(y_0)$ from $0$ to $y_0$ under the projection curve $g(y)$, as shown in Fig. \ref{fig:explained}:
\begin{equation}
\int_{0}^{s_0} \lambda(s') ds' = \int_{0}^{y_0} g(y') dy'
\end{equation}

This problem can be approached by solving the following differential equation:
\begin{equation}\label{eq:ODE}
\begin{cases}
\frac{ds}{dy} = \frac{g(y)}{\lambda(s)}\\
s(0) = 0
 \end{cases}\,.
\end{equation}

Once the transformation $s(y)$ is determined, the screen image can be mapped to the beam's longitudinal phase space. We generate a 2D particle distribution in the ${y-E}$ plane from the screen image by dividing it into vertical slices, each corresponding to a one-dimensional particle density. Using the inverse cumulative distribution function, we create a particle distribution matching the slice density. After establishing a particle distribution equivalent to the screen image, the $y$  coordinates of the particle distribution can be transform to $s$ coordinates. 

\begin{figure}[!htb]
   \centering
   \includegraphics*[width=1\columnwidth]{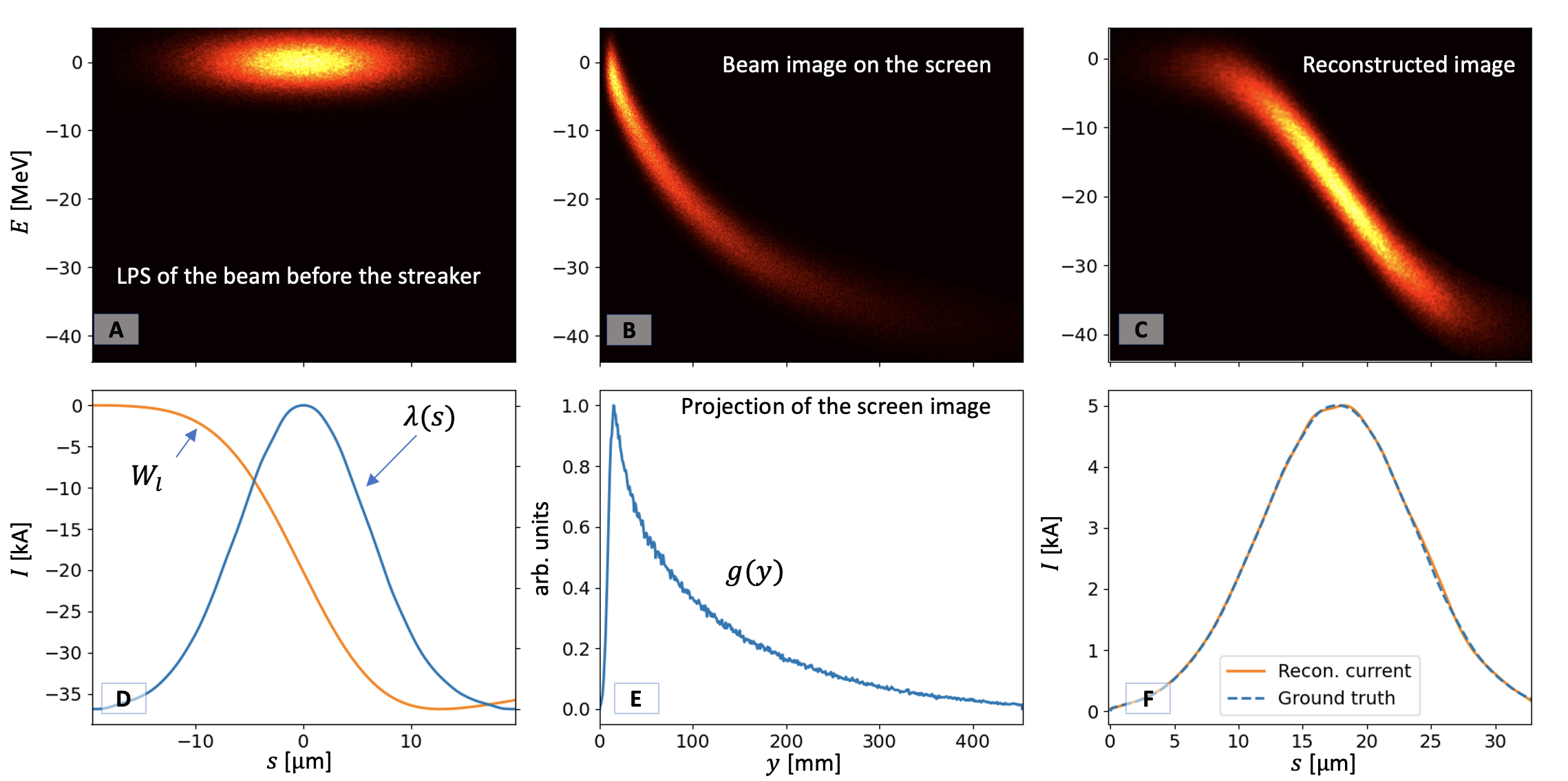}

   \caption{Reconstruction of an ideal Gaussian beam. (A) Initial LPS; (B) Streaked beam image on the screen; (C) Reconstructed LPS; (D) Initial current profile and longitudinal wake \(W_l\); (E) Initial and reconstructed current profiles; (F) Horizontal projection of the streaked image.}

   \label{fig:recon_sim}
\end{figure}

This reconstruction procedure is shown in Fig.~\ref{fig:recon_sim}. A 5 kA, 250 pC Gaussian beam was tracked through the diagnostics beamline with a passive streaker (Fig.~\ref{fig:streaker}). Transverse emittances were set at $\varepsilon_{x,y} = 0.6$ $\mu$m, and the slice energy spread was $\sigma_E = 2.5$ MeV. The beam center was 500 $\mu$m from the corrugated plate. The image projection $g(y)$ and beam current $\lambda(s)$ were used to solve the differential equation Eq.(\ref{eq:ODE}), with LPS reconstruction results shown in Fig.~\ref{fig:recon_sim}C.

As shown in Fig.~\ref{fig:recon_sim}F, the reconstructed current profile closely matches the initial beam distribution. However, the reconstructed longitudinal phase space differs due to the energy chirp induced by longitudinal wakes. For some applications this additional chirp may not be important, and the reconstruction process could conclude at this stage. As a benefit, the reconstruction procedure takes only a fraction of a second and can be used for quasi-real-time LPS observation in the accelerator control room. For more precise energy distribution needs, further improvement of the method will be discussed in the next chapter.

\subsection{Subtraction of induced energy chirp}

Unlike the previous method, identifying the induced energy chirp requires determining the distance between the electron beam and the corrugated plate to calculate and subtract the longitudinal wake \(W_l(s)\) from the reconstructed LPS. The simplest approach involves tracking particles through the beamline while varying this distance and comparing the resulting image with the reference.

We extend the 2D LPS distribution to 6D by adding Gaussian transverse coordinates uniform  in the longitudinal direction. The transverse emittances are set to \(\varepsilon_{x,y} = 0.6\, \mu\text{m}\), matched to the betatron functions at the streaker center. The optimization includes three steps: applying the dipole kick from the streaker, tracking the streaked beam to the screen (excluding second-order effects), and calculating the root mean squared error (RMSE) between the tracked beam image projection, \(g_{tr}^{con}(y)\), and the reference, \(g_{img}(y)\).

\begin{figure}[!htb]
   \centering
   \includegraphics*[width=0.9\columnwidth]{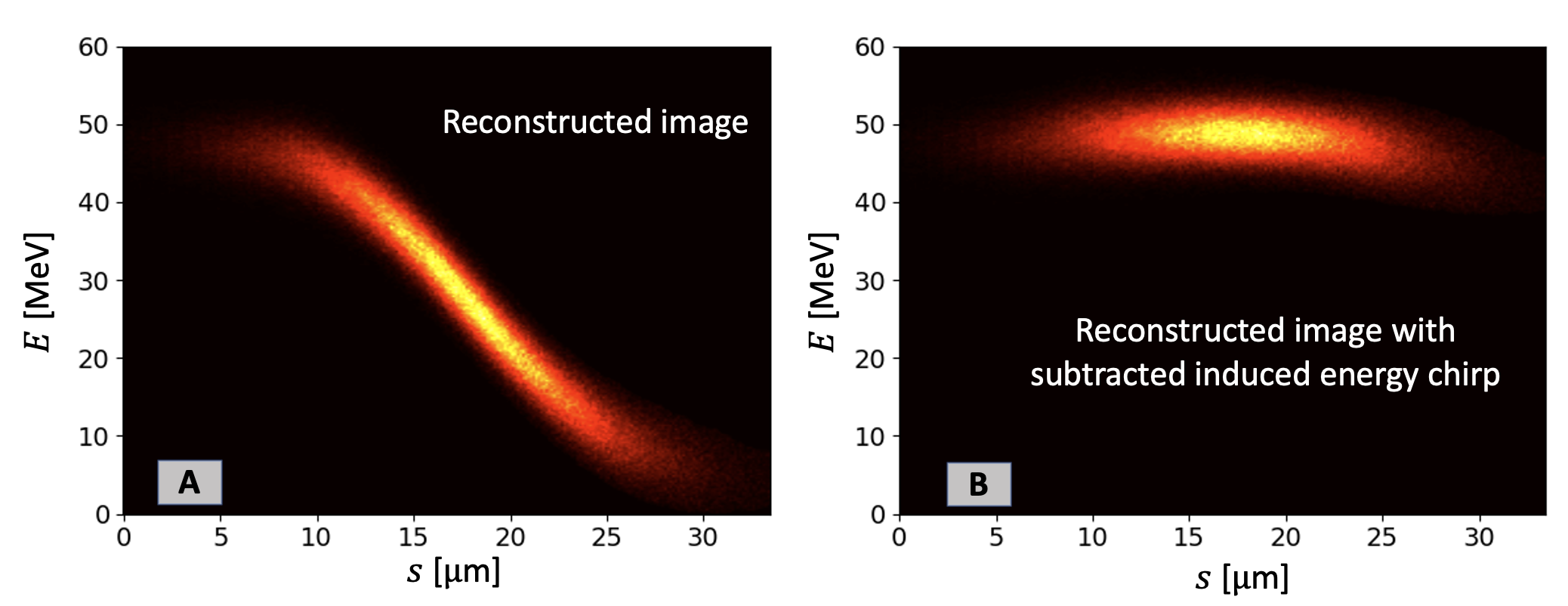}
\caption{Procedure for removing the induced energy chirp by a corrugated structure. (A) Reconstructed LPS; (B) Reconstructed LPS with the induced energy spread subtracted.}

   \label{fig:energy_chirp}
\end{figure}

Using an optimization algorithm, the beam distance to the corrugated structure is determined to be 490 \(\mu\)m, closely matching the reference setting of 500 \(\mu\)m. With the known distance and current profile, the wake potential \(W_l(y)\) is calculated and subtracted from the reconstructed beam distribution, as shown in Fig.~\ref{fig:energy_chirp}. The final reconstructed LPS closely resembles the initial distribution in Fig.~\ref{fig:recon_sim}A, with minor differences due to simplifications like ignoring the quadrupole wake component and slight distance variations.

The optimization procedure takes a fraction of a minute, making it unsuitable for quasi-real-time reconstruction. Therefore, the subtraction of the induced energy chirp is offered as an option for offline reconstruction.

\section{Reconstructing Real-World Measurements}\label{sec5}

\begin{figure}[!htb]
   \centering
   \includegraphics*[width=1\columnwidth]{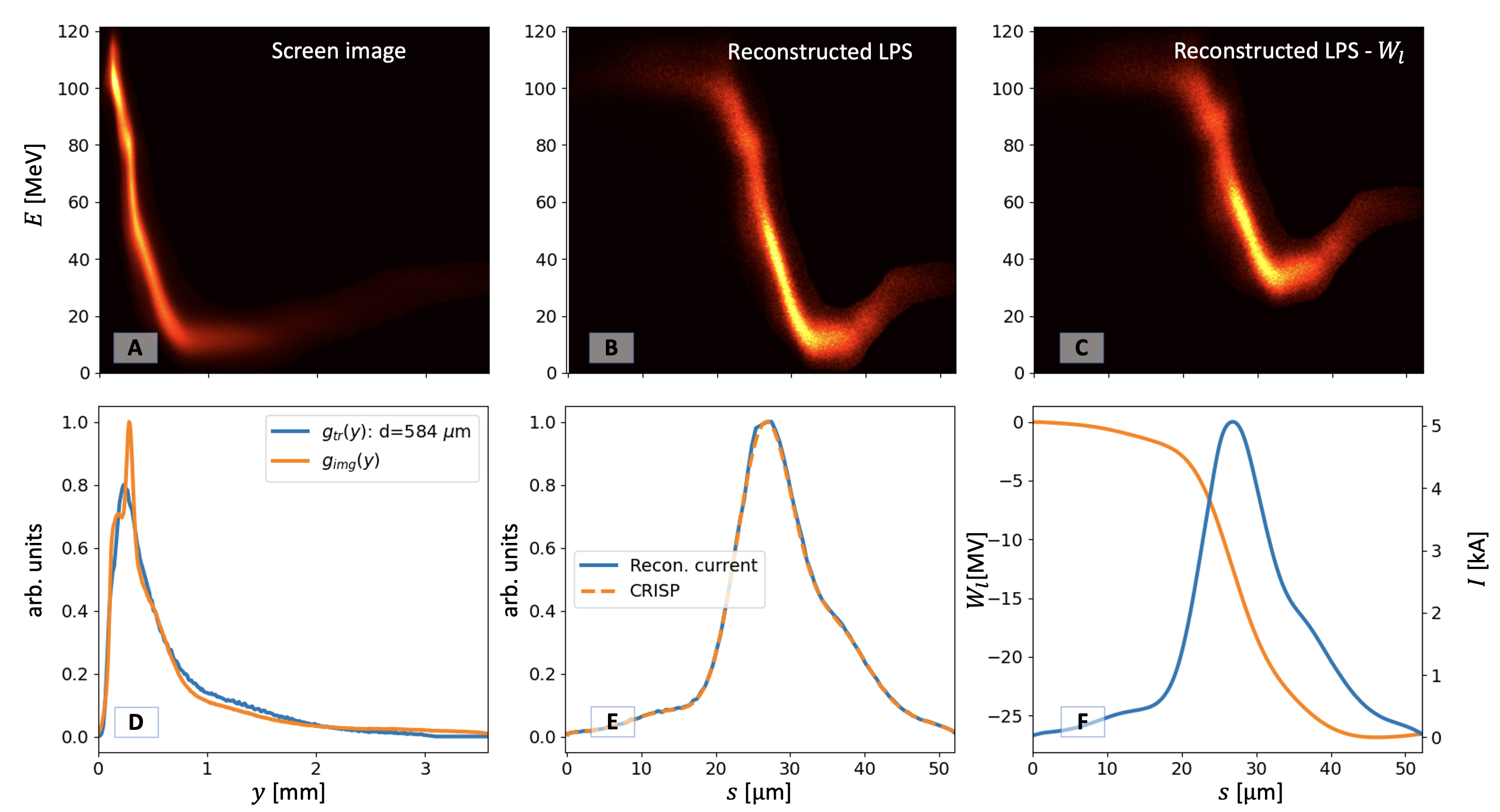}
\caption{Example of the LPS reconstruction with suppressed SASE lasing.   }
   \label{fig:sase_off}
\end{figure}

\begin{figure}[!htb]
   \centering
   \includegraphics*[width=1\columnwidth]{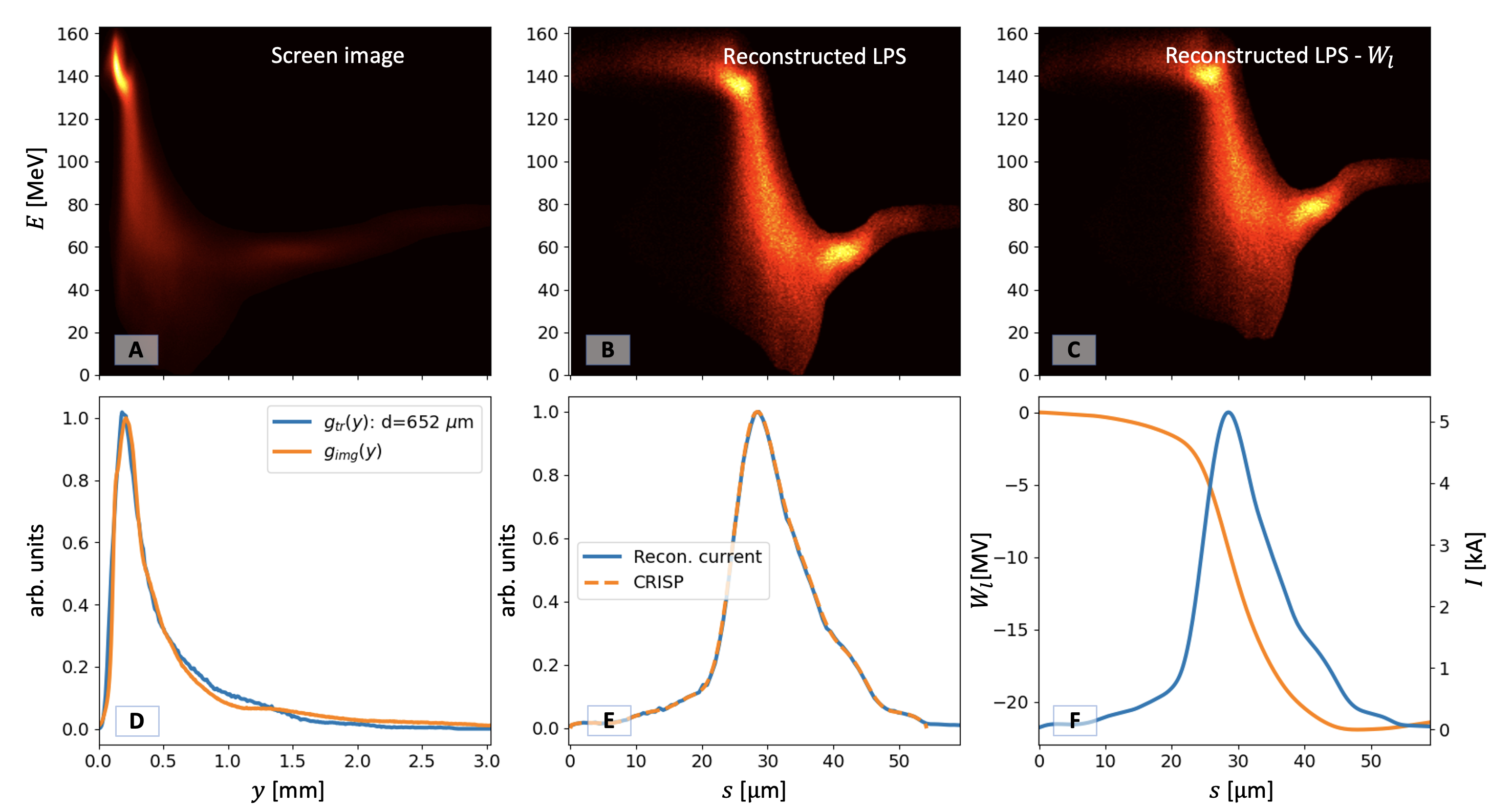}
\caption{Example of the LPS reconstruction with SASE lasing.   }
   \label{fig:sase_on}
\end{figure}

We applied the reconstruction method to LPS measurements in two scenarios: first, after the SASE2 undulator with SASE lasing suppressed by a transverse kick (Fig.~\ref{fig:sase_off}); and second, after the SASE2 undulator with lasing at 1.6 mJ and 9 keV photon energy (Fig.~\ref{fig:sase_on}).


\section{Discussion}\label{sec4}

Real-world measurements show that the reconstructed current profiles closely match the CRISP profiles in both cases (Fig.~\ref{fig:sase_off}E, Fig.~\ref{fig:sase_on}E), highlighting the importance of accurate beam current profiles for reconstruction. However, CRISP has limitations: its 8 fs resolution may miss high spikes, lower current amplitudes lead to noisy diffraction radiation spectra, and it cannot distinguish the beam's head from the tail. Potential solutions include enhancing CRISP with machine learning \cite{undul_taper} or using a TDS after the last bunch compressor for noninvasive current profile measurements but requires the development of server-based TDS measurements, which are already in the planning stages.

%
%
%

%
%



\begin{thebibliography}{9} 
	
\bibitem{TDS}
P. Emma, J. Frisch,  P. Krejcik,  A transverse rf deflecting structure for bunch length and phase space diagnostics. Tech. Rep. LCLS- TN-00-12, SLAC, https://www-ssrl.slac.stanford.edu/lcls/technotes/lcls-tn-00-12.pdf (2000)

\bibitem{XTCAV2}
C. Behrens, F.-J. Decker, Y. Ding, V. A. Dolgashev, J. Frisch, Z. Huang, P. Krejcik, H. Loos, A. Lutman, T. J. Maxwell, J. Turner, J. Wang, M.-H. Wang, J. Welch, and J. Wu, Few-femtosecond time-resolved measurements of x-ray freeelectron lasers, Nat. Commun. 5, 3762 (2014).


\bibitem{XFEL}
W. Decking, S. Abeghyan, P. Abramian, A. Abramsky, A. Aguirre, C. Albrecht, P. Alou, M. Altarelli, P. Altmann, K. Amyan, V. Anashin, E. Apostolov, K. Appel, D. Auguste, V. Ayvazyan, S. Baark, F. Babies, N. Baboi, P. Bak, V. Balandin et al., A MHz-repetition-rate hard x-ray free-electron laser driven by a superconducting linear accelerator, Nat. Photonics 14, 391 (2020).




\bibitem{Bane_2012} 
K.L.F. Bane and G. Stupakov. Corrugated pipe as a beam dechirper. Nuclear Instruments and Methods in Physics Research
Section A: Accelerators, Spectrometers, Detectors and Associated Equipment, 690:106 – 110, 2012.

\bibitem{Emma2014}
P. Emma, M. Venturini, K. L. F. Bane, G. Stupakov, H.-S. Kang, M. S. Chae, J. Hong, C.-K. Min, H. Yang, T. Ha et al., Experimental Demonstration of Energy-Chirp Control in Relativistic Electron Bunches Using a Corrugated Pipe, Phys. Rev. Lett. 112, 034801 (2014).

\bibitem{novo2015}
A. Novokhatski, Wakefield potentials of corrugated structures, Phys. Rev. Spec. Top. Accel. Beams 18, 104402 (2015).

\bibitem{bane_zagor2016}
K. Bane, G. Stupakov, and I. Zagorodnov, Analytical formulas for short bunch wakes in a flat dechirper, Phys. Rev. Accel. Beams 19, 084401 (2016).



\bibitem{Bettoni2016}
S. Bettoni, P. Craievich, A. A. Lutman, and M. Pedrozzi, Temporal profile measurements of relativistic electron bunch based on wakefield generation, Phys. Rev. Accel. Beams 19, 021304 (2016).

\bibitem{Seok2018}
J. Seok, M. Chung, H.-S. Kang, C.-K. Min, and D. Na, Use of a corrugated beam pipe as a passive deflector for bunch length measurements, Phys. Rev. Accel. Beams 21, 022801 (2018).

\bibitem{PSI2022}
Philipp Dijkstal, Alexander Malyzhenkov, Paolo Craievich , Eugenio Ferrari, Romain Ganter, Sven Reiche, Thomas Schietinger, Pavle Jurani', and Eduard Prat,  Self-synchronized and cost-effective time-resolved measurements at x-ray free-electron lasers with femtosecond resolution, Phys. Rev. Res. 4, 013017 (2022).







\bibitem{tomin_ipac}
S. Tomin, W. Decking, N. Golubeva, A. Novokshonov, T. Wohlenberg, and I. Zagorodnov. “Longitudinal Phase Space Diagnostics with Corrugated Structure at the European XFEL.” en. In: Proc. IPAC 2022, Bangkok, Thailand. JACoW Publishing, Geneva, Switzerland, 2022, pp. 275–278. doi:10.18429/JACOW-IPAC2022-MOPOPT020.

\bibitem{dijkstal2024}
P. Dijkstal, W. Qin, and S. Tomin, Longitudinal phase space measurements with a passive streaker at the European XFEL, PRAB


\bibitem{CRISPwesch}
    S. Wesch, B. Schmidt, C. Behrens, H. Delsim-Hashemi, P. Schmüser, A multi-channel THz and infrared spectrometer for femtosecond electron bunch diagnostics by single-shot spectroscopy of coherent radiation, Nucl. Instrum. Methods Phys. Res., Sect. A 665 40 (2011).

\bibitem{CRISP}
	N. M. Lockmann, C. Gerth, B. Schmidt, S. Wesch, Noninvasive Thz Spectroscopy for bunch Current Profile Reconstructions at Mhz Repetition Rates. Phys Rev Accel Beams (2020) 23:112801. doi:10.1103/PhysRevAccelBeams.23.112801

\bibitem{CRISPrec}
B. Schmidt, N. M. Lockmann, P. Schmüser, and S. Wesch, Benchmarking Coherent radiation spectroscopy as a tool for high-resolution bunch shape reconstruction at free-electron lasers, Phys. Rev. Accel. Beams 23, 062801 (2020).




\bibitem{undul_taper}
S. Tomin, J. Kaiser, N. Lockmann, T. Wohlenberg, I. Zagorodnov, Undulator linear taper control at the European X-Ray Free-Electron Laser facility, Phys. Rev. Accel. Beams 27, 042801 (2024)



 

	\end{thebibliography}
\end{document}